\documentclass{emulateapj}

\usepackage{natbib}
\usepackage{amsmath}

\newcommand{\gri}{\protect\hbox{$gri$} }
\newcommand{\ri}{\protect\hbox{$ri$} }

\newcommand{\ubvrr}{\protect\hbox{$U\!BV\!r$} }
\newcommand{\ubvri}{\protect\hbox{$U\!BV\!ri$} }
\newcommand{\ubv}{\protect\hbox{$U\!BV$} }

\newcommand{\about}{$\sim\!\!$~}
\newcommand{\kms}{\,km\,s$^{-1}$}
\newcommand{\sn}{SN~2006bt}
\newcommand{\host}{CGCG~108-013}

\newcommand{\err}[2]{\ensuremath{^{+#1}_{-#2}}}

\def\lsim{\hbox{\rlap{\raise 0.425ex\hbox{$<$}}\lower 0.65ex\hbox{$\sim$}}}
\def\gsim{\hbox{\rlap{\raise 0.425ex\hbox{$>$}}\lower 0.65ex\hbox{$\sim$}}}
\def\arcmin{\hbox{$^\prime$}}
\def\arcsec{\hbox{$^{\prime\prime}$}}
\def\arcdeg{\mbox{$^\circ$}}

\shorttitle{The Troublesome \sn}
\shortauthors{Foley et~al.}

\begin{document}

 \title{\sn: A Perplexing, Troublesome, and Possibly Misleading Type~Ia Supernova}

\def\cfa{1}
\def\clay{2}
\def\harv{3}
\def\berk{4}

\author{
{Ryan~J.~Foley}\altaffilmark{\cfa,\clay}, 
{Gautham~Narayan}\altaffilmark{\cfa,\harv},
{Peter~J.~Challis}\altaffilmark{\cfa},
{Alexei~V.~Filippenko}\altaffilmark{\berk},
{Robert~P.~Kirshner}\altaffilmark{\cfa},
{Jeffrey~M.~Silverman}\altaffilmark{\berk}, and
{Thea~N.~Steele}\altaffilmark{\berk}
}

\altaffiltext{\cfa}{
Harvard-Smithsonian Center for Astrophysics,
60 Garden Street, 
Cambridge, MA 02138.
}
\altaffiltext{\clay}{
Clay Fellow. Electronic address rfoley@cfa.harvard.edu .
}
\altaffiltext{\harv}{
Department of Physics,
Harvard University,
17 Oxford Street,
Cambridge, MA 02138.
}
\altaffiltext{\berk}{
Department of Astronomy,
University of California,
Berkeley, CA 94720-3411.
}

\begin{abstract}
\sn\ displays characteristics unlike those of any other known Type Ia
supernova (SN~Ia).  We present optical light curves and spectra of
\sn\ which demonstrate the peculiar nature of this object.  \sn\ has
broad, slowly declining light curves indicative of a hot,
high-luminosity SN, but lacks a prominent second maximum in the $i$
band as do low-luminosity SNe~Ia.  Its spectra are similar to those of
low-luminosity SNe~Ia, containing features that are only present in
cool SN photospheres.  Light-curve fitting methods suggest that \sn\
is reddened by a significant amount of dust; however, it occurred in
the outskirts of its early-type host galaxy and has no strong Na~D
absorption in any of its spectra, suggesting a negligible amount of
host-galaxy dust absorption.  \ion{C}{2} is possibly detected in our
pre-maximum spectra, but at a much lower velocity than other elements.
The progenitor was likely very old, being a member of the halo
population of a galaxy that shows no signs of recent star formation.
SNe~Ia have been very successfully modeled as a one-parameter family,
and this is fundamental to their use as cosmological distance
indicators.  \sn\ is a challenge to that picture, yet its relatively
normal light curves allowed \sn\ to be included in cosmological
analyses.  We generate mock SN~Ia datasets which indicate that
contamination by similar objects will both increase the scatter of a
SN~Ia Hubble diagram and systematically bias measurements of
cosmological parameters.  However, spectra and rest-frame $i$-band
light curves should provide a definitive way to identify and eliminate
such objects.
\end{abstract}

\keywords{supernovae: general --- supernovae: individual (\sn)}

\defcitealias{Hicken09:lc}{H09a}
\defcitealias{Hicken09:de}{H09b}


\section{Introduction}\label{s:intro}

A general model of Type Ia supernova (SN~Ia) explosions was
established decades ago \citep[e.g.,][]{Whelan73, Nomoto84:w7,
Colgate69, Arnett82}.  In this model, a CO white dwarf accretes
material from a companion star until it reaches the density for carbon
ignition, causing a runaway nuclear reaction to propagate through the
star.  During this process, \about $0.6 {\rm M}_{\sun}$ of the star
burns to $^{56}$Ni which powers the luminosity of the SN by decaying
to $^{56}$Co and then to $^{56}$Fe \citep[e.g.,][]{Colgate69}.

This scenario predicts a close correspondence between the amount of
$^{56}$Ni, the peak luminosity of the SN, and the temperature of the
ejecta which determines the ionization of the ejecta \citep{Arnett82}.
These predictions have been verified by observations of normal SNe~Ia
as well as both high-luminosity \citep[e.g.,
SN~1991T;][]{Filippenko92:91T, Phillips92} and low-luminosity
\citep[e.g., SN~1991bg;][]{Filippenko92:91bg, Leibundgut93} SNe~Ia.
These relationships predict a spectral sequence from cool to hot
photospheres \citep{Hachinger08} which have also been empirically
determined \citep[e.g.,][]{Nugent95}.  The width-luminosity
relationship (WLR) between light-curve shape and luminosity that
empowers cosmologists to use SNe~Ia as accurate and precise distance
indicators \citep{Phillips93} is directly related to the ionization
(and therefore the temperature) of the SN photosphere, specifically
when doubly ionized Fe-group elements recombine to singly ionized ions
\citep{Kasen07:wlr}.

Except for a few examples of very peculiar SNe~Ia such as SN~2000cx
\citep{Li01:00cx}, the SN~2002cx-like class of objects
\citep{Li03:02cx, Phillips07, Foley09:08ha}, hydrogen-rich objects
\citep{Hamuy03, Aldering06}, and potential super-Chandrasekhar objects
\citep{Howell06, Hicken07}, SNe~Ia are observationally well described
by their light-curve shape alone.  There are methods of further
subclassifying objects based on spectral characteristics
\citep{Benetti05, Branch06}.  \citet{Foley08:uv} and
\citet{Wang09:2pop} showed that incorporating spectral features into
distance measurements from light-curve fitting can improve SN distance
measurements, but the former study was based on a small sample and it
is unclear if the improvement seen in the latter study is intrinsic to
the explosions or related to the SN environment.  Most physical
parameters and observables of SNe~Ia, such as $^{56}$Ni mass, peak
luminosity, light-curve shape, photospheric temperature, and ejecta
ionization, can be parameterized by a single number (excluding clear
outliers like the objects mentioned above); hence, knowing any one of
the parameters or observables will tell you the rest.

\sn\ was discovered by \citet{Lee06} on 2006 April 26.5 (UT dates
will be used throughout this paper) at (R.A., Dec.) = (15:56:30.526,
+20:02:45.34) (J2000), 44.4\arcsec\ west and 22.9\arcsec\ south of
\host, the presumed host galaxy.  \citet{Filippenko06:06bt} obtained a
spectrum one night after discovery and determined that \sn\ was a
SN~Ia with some slight peculiarities.  Photometric observations of the
SN were then performed using the 1.2~m telescope located at the Fred
L.\ Whipple Observatory (FLWO), and light curves derived from these
observations were published by \citet{Hicken09:lc} (hereafter H09a).
Fitting these light curves, \citet{Hicken09:de} (hereafter H09b)
determined that \sn\ peaked in the $B$ band on 2006 May 1.4, with
$\Delta m_{15} (B) = 1.09 \pm 0.06$~mag and $\Delta m_{15} (V) = 0.54
\pm 0.04$~mag, indicating that it had a slightly broad light curve and
should thus have a slightly higher than average peak luminosity
\citep{Phillips93}.  Fitting the light curves with more sophisticated
techniques, they found the SALT2 \citep{Guy07} and MLCS2k2 \citep[with
$R_{V} = 1.7$;][]{Jha07} fit parameters of $x1 = 0.069 \pm 0.104$, $c
= 0.162 \pm 0.008$ and $\Delta = -0.325 \pm 0.052$, $A_{V} = 0.428 \pm
0.053$ mag, respectively.  All light-curve fits produced reasonable
values of $\chi^{2}$ and there was no obvious reason to suspect any
peculiarity from quality cuts.  Both SALT2 and MLCS2k2 indicate that
\sn\ had a high luminosity and a red color (which MLCS2k2 interprets
as a significant amount of host-galaxy extinction, but SALT2 does not
attempt to distinguish between intrinsic color differences and
reddening from dust).  Using these light-curve fits, \sn\ was included
in a cosmological analysis of a large set of SN data
\citepalias{Hicken09:de}.  \sn\ was not an outlier in the Hubble
diagram and its inclusion in the analysis did not significantly change
the measured cosmological parameters.

We present our data on \sn\ in Section~\ref{s:obs}, performing a
detailed analysis of the photometry and spectroscopy in
Sections~\ref{s:phot} and \ref{s:spec}, respectively.  In
Section~\ref{s:enviro}, we investigate the SN environment.  We discuss
possible physical interpretations of \sn\ as well as implications for
SN~Ia cosmology in Section~\ref{s:disc}.  Our results are summarized
in Section~\ref{s:conc}.


\section{Observations and Data Reduction}\label{s:obs}

\begin{figure}
\begin{center}
\epsscale{1.5}
\rotatebox{90}{
\plotone{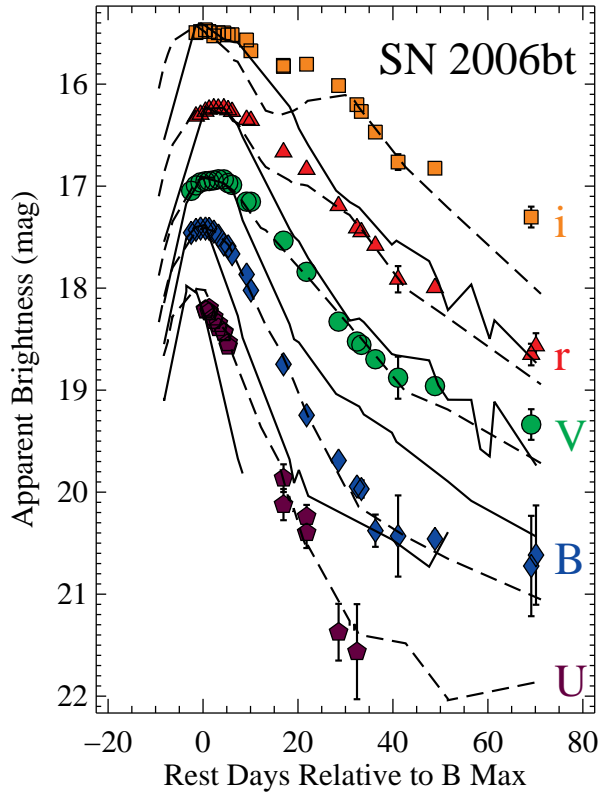}}
\caption{\ubvri\ light curves of \sn\ shifted by 1.2, 0.27, 0,
$-0.73$, and $-1.8$~mag, respectively.  Also plotted are the light
curves of SNe~2005mz ($\Delta m_{15} (B) = 1.96$~mag; solid lines) and
2006ax ($\Delta m_{15} (B) = 1.08$~mag; dashed lines).  The light
curves of SNe~2005mz and 2006ax have been shifted to match the peak of
\sn\ in each band.  The \ubv light curves of \sn\ are very similar to
those of SN~2006ax, while the $r$ and $i$ bands show significant
differences near the ``shoulder'' in the $r$ band and the ``trough''
and second maximum in the $i$ band.  [{\it See the electronic edition
of the Journal for a color version of this figure.}]}\label{f:lc}
\end{center}
\end{figure}

\sn\ was followed photometrically by the 1.2~m telescope located at
FLWO and its \ubvri\ light curves were originally presented by
\citetalias{Hicken09:lc}.  We reproduce those light curves in
Figure~\ref{f:lc}.  Because our conclusions are extremely sensitive to
the nature of the light curves, we have re-reduced the data and
compared our photometry of the field stars used for photometric
calibration against photometry of the stars from SDSS in \ri\ and
$V$\footnote{http://www.sdss.org/dr7/algorithms/ \\
sdssUBVRITransform.html\#Lupton2005}; all measurements were found to
be within 0.01~mag of the photometry published by
\citetalias{Hicken09:lc}.

\begin{figure}
\begin{center}
\epsscale{1.15}
\rotatebox{0}{
\plotone{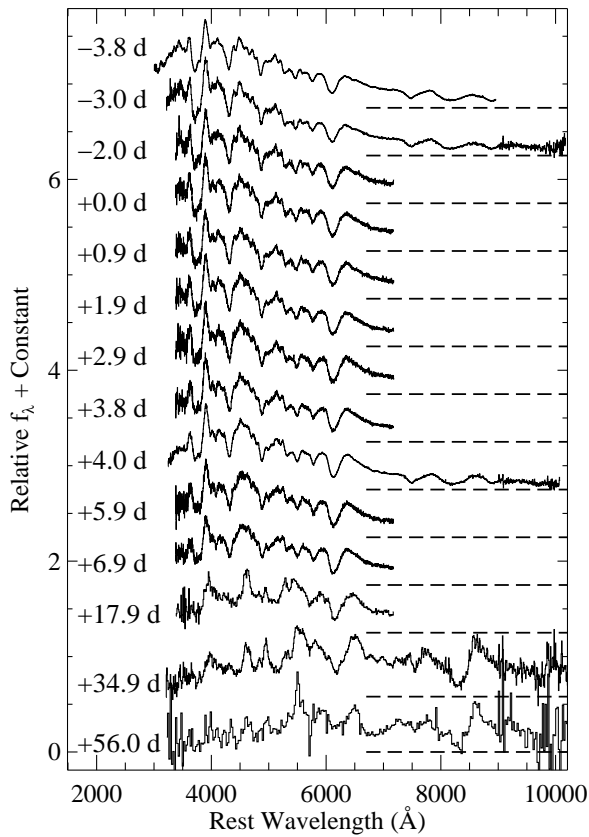}}
\caption{Optical spectra of \sn.  The phase relative to $B$ maximum is
noted to the left of each spectrum.  The zero flux level for each
spectrum is indicate by the dashed line.}\label{f:spec}
\end{center}
\end{figure}

We also obtained several low-resolution spectra of \sn\ with the FAST
spectrograph \citep{Fabricant98} on the FLWO 1.5~m telescope (as part
of the CfA SN monitoring program; Blondin et~al., in prep.), the Kast
double spectrograph \citep{Miller93} on the Shane 3~m telescope at
Lick Observatory, and the Low Resolution Imaging Spectrometer
\citep[LRIS;][]{Oke95} on the 10~m Keck~I telescope; a journal of
observations can be found in Table~\ref{t:spec}.  Standard CCD
processing and spectrum extraction were performed with
IRAF\footnote{IRAF: the Image Reduction and Analysis Facility is
distributed by the National Optical Astronomy Observatory, which is
operated by the Association of Universities for Research in Astronomy,
Inc. (AURA) under cooperative agreement with the National Science
Foundation (NSF).}.  The data were extracted using the optimal
algorithm of \citet{Horne86}.  Low-order polynomial fits to
calibration-lamp spectra were used to establish the wavelength scale,
and small adjustments derived from night-sky lines in the object
frames were applied.  We employed our own IDL routines to flux
calibrate the data and remove telluric lines using the well-exposed
continua of spectrophotometric standard stars \citep{Wade88, Foley03}.
Our spectra of \sn\ are presented in Figure~\ref{f:spec}.

\begin{deluxetable*}{lllrl}
\tabletypesize{\scriptsize}
\tablewidth{0pt}
\tablecaption{Log of Spectral Observations\label{t:spec}}
\tablehead{
\colhead{} &
\colhead{} &
\colhead{Telescope /} &
\colhead{Exposure} &
\colhead{} \\
\colhead{Phase\tablenotemark{a}} &
\colhead{UT Date} &
\colhead{Instrument} &
\colhead{(s)} &
\colhead{Observer\tablenotemark{b}}}

\startdata

$-3.8$ & 2006 Apr.\ 27.6 & Keck~I/LRIS      &  300                 & AF, RF \\
$-3.0$ & 2006 Apr.\ 28.4 & Lick/Kast        & 1800                 & DW, JS \\
$-2.0$ & 2006 Apr.\ 29.4 & FLWO/FAST        & 1200                 & WB \\
   0.0 & 2006 May    1.4 & FLWO/FAST        & 1500                 & WB \\
   0.9 & 2006 May    2.3 & FLWO/FAST        & 1500                 & WB \\
   1.9 & 2006 May    3.3 & FLWO/FAST        & 1500                 & TG \\
   2.9 & 2006 May    4.3 & FLWO/FAST        & 1800                 & TG \\
   3.9 & 2006 May    5.3 & FLWO/FAST        & 1500                 & WP \\
   4.0 & 2006 May    5.4 & Lick/Kast        & 1800                 & JS, MM, RF \\
   5.9 & 2006 May    7.3 & FLWO/FAST        & 1500                 & WP \\
   6.9 & 2006 May    8.3 & FLWO/FAST        & 1500                 & WP \\
  17.9 & 2006 May   19.3 & FLWO/FAST        & 1800                 & MC \\
  34.9 & 2006 Jun.\  5.4 & Lick/Kast        & 2100                 & JS, MM \\
  56.0 & 2006 Jun.\ 26.4 & Lick/Kast        & 2400                 & DW, JS, TS

\enddata

\tablenotetext{a}{Days since $B$ maximum, 2006 May 1.4 (JD 2,453,856.9).}

\tablenotetext{b}{AF = A.\ Filippenko, DW = D.\ Wong, JS = J.\
Silverman, MC = M.\ Calkins, MM = M.\ Moore, RF = R.\ Foley, TG = T.\
Groner, TS = T.\ Steele, WB = W.\ Brown, WP = W.\ Peters.}

\end{deluxetable*}


\section{Photometric Analysis}\label{s:phot}

\subsection{Light Curves}

The light curves of \sn\ have already been published by
\citetalias{Hicken09:lc}.  The light curves were fit by both
\citetalias{Hicken09:lc} and \citetalias{Hicken09:de}, finding $\Delta
m_{15} (B) = 1.09 \pm 0.06$~mag and $\Delta m_{15} (V) = 0.54 \pm
0.04$~mag.  They similarly found the SALT2 \citep{Guy07} and MLCS2k2
\citep[with $R_{V} = 1.7$);][]{Jha07} fit parameters of $x1 = 0.069
\pm 0.104$, $c = 0.162 \pm 0.008$ and $\Delta = -0.325 \pm 0.052$,
$A_{V} = 0.428 \pm 0.053$ mag, respectively.  The light-curve fits
indicate that \sn\ has a slightly slow decline, and a red color
(SALT2) or relatively high extinction (MLCS2k2). The MLCS2k2 fit has a
reduced $\chi^{2}$ of \about 1.36 for 98 separate observations,
including pre-maximum data in multiple passbands.  The SALT2 templates
used by \citetalias{Hicken09:lc} do not cover the late-time photometry
and the best fit has a reduced $\chi^{2}$ of \about 2 with 70
observations. However, estimates of the time of maximum and the
best-fit light-curve models using either method are quite comparable.
An analysis of the light-curve fits alone would lead one to conclude
that \sn\ is not a particularly discrepant object and can be included
in any cosmological analysis.

However, the host galaxy, \host, is an S0/a galaxy (with $d = 139
Mpc$) and \sn\ has a projected galactocentric distance (PGCD) of
33.7~kpc (equivalent to $7.8 R_{\rm eff}$, where $R_{\rm eff} =
6.4\arcsec$ is defined by the Petrosian radius where 50\% of the flux
is contained within a circle of radius $R_{\rm eff}$; see
Section~\ref{s:enviro}), making \sn\ have the fourth largest PGCD of
SNe~Ia in the CfA3 sample of 185 SNe~Ia. The SN environment is thus
expected to have little dust (excluding possible circumstellar dust).
Additionally, \citetalias{Hicken09:de} shows the relationship between
host morphology, PGCD, and light-curve fit parameters.  In the CfA3
sample, \sn\ has among the largest values for $x1$, $c$, and $A_{V}$,
and the smallest values of $\Delta$ for all SNe hosted in early-type
galaxies.  Furthermore, \sn\ is a clear outlier in the trend of
smaller $c$ or $A_{V}$ with increasing PGCD (see Figures~18 and 19 in
\citetalias{Hicken09:de}).

From the light curves, we can see that \sn\ evolves slightly
differently from other SNe with similar $B$-band decline rates.  In
Figure~\ref{f:lc}, the light curves of \sn\ are compared to those of
SN~2006ax, an object with $\Delta m_{15} (B) = 1.08 \pm 0.05$~mag and
minimal ($A_{V} < 0.04$~mag) host-galaxy extinction.  These objects
have very similar \ubv\ light curves; however, the $r$ and $i$ light
curves have some slight differences.  In the $r$ band, \sn\ lacks the
``shoulder'' at 15--20~days after maximum brightness seen in average
and high-luminosity SNe~Ia.  In the $i$ band, \sn\ has a shoulder, but
lacks the deep ``trough'' and subsequent prominent second maximum seen
in most SNe~Ia.

In Figure~\ref{f:lc}, the light curves of \sn\ are also compared to
those of SN~2005mz, a fast-declining SN~Ia ($\Delta m_{15} (B) = 1.96
\pm 0.14$~mag) with moderate host-galaxy extinction ($A_{V} = 0.27 \pm
0.09$~mag).  Although this object lacks a shoulder in either the $r$
or $i$ bands, it declines much faster in all bands.

\subsection{Color Curves}

\begin{figure}
\begin{center}
\epsscale{1.5}
\rotatebox{90}{
\plotone{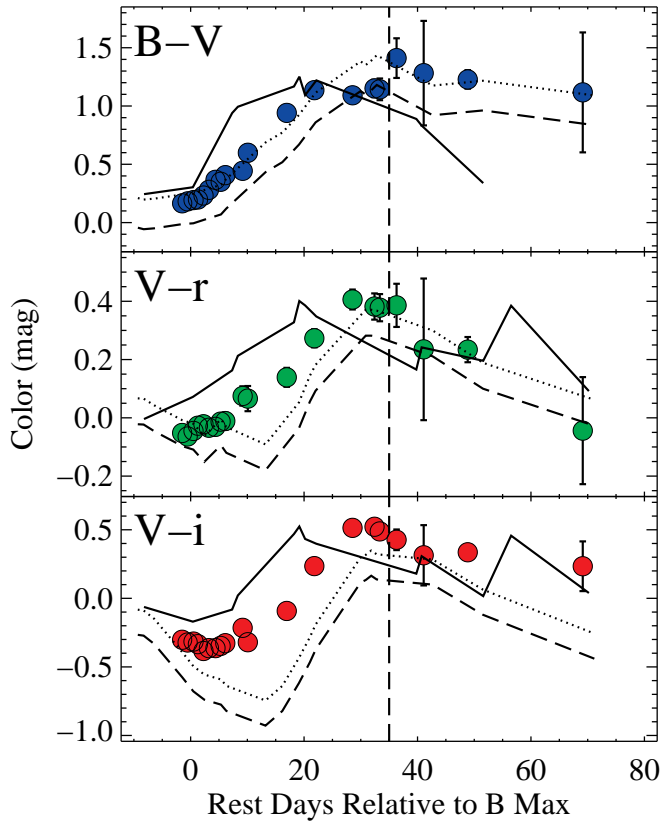}}
\caption{$B-V$, $V-r$, and $V-i$ color curves of \sn.  The color
curves of SNe~2005mz ($\Delta m_{15} (B) = 1.96$~mag; corrected for
$A_{V} = 0.27$~mag; solid lines) and 2006ax ($\Delta m_{15} (B) =
1.08$~mag; dashed lines) are overplotted.  We also plot the color
curves of SN~2006ax reddened by $A_{V} = 0.428$~mag (with $R_{V} =
1.7$; dotted lines) to match the measured reddening determined by the
MLCS2k2 fit to \sn\ \citepalias{Hicken09:lc}.  The color curves are
very different from those of both a normal SN~Ia with a similar
decline rate (SN~2006ax) and a low-luminosity SN~Ia with a similar
maximum-light spectrum (SN~2005mz).  Furthermore, although the
reddened $B-V$ color curve of the normal SN~Ia is similar to that of
\sn, the $V-r$ and $V-i$ color curves are quite different in the
ranges $10 < t < 25$~days and $0 < t < 30$~days relative to maximum,
respectively.  It is worth noting that after the extinction
correction, \sn\ has a similar color to a normal SN~Ia at $t =
35$~days after maximum brightness, when the colors of unreddened
SNe~Ia are very similar regardless of decline rate \citep{Lira98}.
[{\it See the electronic edition of the Journal for a color version of
this figure.}]}\label{f:cc}
\end{center}
\end{figure}

Figure~\ref{f:cc} presents the color curves of SNe~2005mz, 2006ax, and
2006bt We have corrected the color curves of SN~2005mz for host-galaxy
extinction as determined by MLCS fits ($A_{V} = 0.27$~mag and $R_{V} =
1.7$), and all objects for Milky Way extinction as determined by
\citet{Schlegel98}, which indicates $E(B-V) = 0.051$~mag for \sn.  The
curves of \sn\ generally fall between the curves of SNe~2005mz and
2006ax for $t < 35$~days and above both curves for $t > 35$~days.
Reddening the color curves of SN~2006ax by $A_{V} = 0.428$~mag with $R_{V}
= 1.7$, the best-fit value for \sn\ from \citetalias{Hicken09:de}
using MLCS, the $B-V$ color curves of SNe~2006ax and 2006bt are very
well matched.  This is consistent with the ``Lira relation''
\citep{Phillips99}, which uses the late-time colors of SNe~Ia to
predict their host-galaxy reddening.  However, this same correction
does not account for the differences in the $V-r$ and $V-i$ color
curves of SNe~2006ax and 2006bt.  Given the shapes of their color
curves, reddening alone cannot account for the differences between
these objects.  Furthermore, the color curves of SNe~2005mz and 2006bt
differ by more than reddening alone.

Considering the PGCD, the early-type host, the non-standard color
curves, and a lack of Na~D absorption in any spectrum (see
Section~\ref{s:spec}), we believe the MLCS-fit extinction of $A_{V} =
0.428$~mag to be unrepresentative of the true host-galaxy extinction.
Given these factors, the extinction is most likely very small, and for
the remainder of our analysis, we assume that there is no host-galaxy
extinction.

\subsection{Luminosity}

Since \sn\ has slightly odd light and color curves as well as a
host-galaxy extinction very different from its MLCS fit extinction
(which indicates that \sn\ is not well modeled by the training set of
MLCS), it is worth investigating the possibility that \sn\ does not
follow the WLR.  The MLCS model would indicate that \sn\ had a peak
absolute magnitude in $V$ that is 0.428~mag (equivalent to the fit
value for $A_{V}$) brighter than its true value.

\begin{figure}
\begin{center}
\epsscale{1.1}
\rotatebox{90}{
\plotone{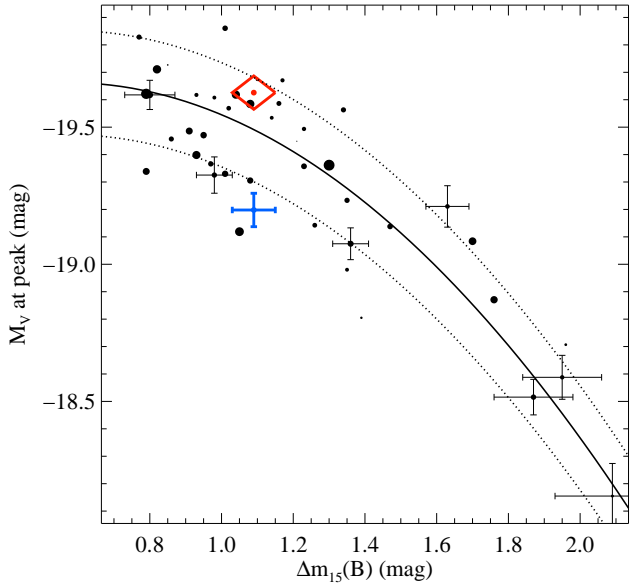}}
\caption{Relationship between peak absolute magnitude ($M_{V}$) and
decline rate ($\Delta m_{15} (B)$) for a subsample of the CfA3 dataset
with $A_{V} < 1$~mag, $E(B-V)_{\rm MW} < 0.5$~mag, $z > 0.015$, and no
major spectroscopic or photometric peculiarities (for example,
SN~2002cx was excluded).  To be consistent with
\citetalias{Hicken09:de}, we set $H_{0} = 65$~km~s$^{-1}$~Mpc$^{-1}$.
The size of each point is inversely proportional to its uncertainties.
A few points have uncertainties plotted to show the relationship.
Overplotted is the quadratic relationship between $\Delta m_{15} (B)$
and $M_{V}$ given by \citet{Phillips99} ($M_{V} (\Delta m_{15} =
1.1~{\rm mag}) = -19.48$~mag; solid lines) as well as that
relationship shifted by an amount corresponding to the intrinsic
scatter in the relationship (dotted lines).  \sn\ is plotted both with
its true value for $M_{V}$ (blue point with solid error bars) and
corrected by the MLCS fit value for $A_{V}$ (red point with
uncertainties indicated by the vertices of the surrounding diamond).
Both points are consistent with the relationship.  [{\it See the
electronic edition of the Journal for a color version of this
figure.}]}\label{f:phillips}
\end{center}
\end{figure}

We plot the relationship between $\Delta m_{15} (B)$ and $M_{V}$ in
Figure~\ref{f:phillips} for a subsample of the CfA3 sample
\citepalias{Hicken09:lc}.  Using the extinction derived from MLCS,
\sn\ is slightly overluminous relative to the relationship fit by
\citet{Phillips99}, but it is within the scatter around the
relationship.  With no extinction correction, \sn\ is slightly
underluminous for its light-curve shape; however, it is not a
significant outlier.

Despite an atypical set of light curves, \sn\ is not a significant
outlier to the WLR using either no extinction or the MLCS-determined
extinction value.  For SN~2006bt, \citetalias{Hicken09:de} measured
distance moduli of $\mu = 35.981$, 35.944, 35.784, and 35.993~mag with
distance fitters SALT, SALT2, MLCS with $R_{V} = 3.1$, and MLCS with
$R_{V} = 1.7$, respectively.  Using the CMB-corrected redshift for
\host\ of $z_{\rm CMB} = 0.0330$ and $H_{0} =
65$~km~s$^{-1}$~Mpc$^{-1}$ (to match the analysis of
\citetalias{Hicken09:de}), \host\ has a Hubble-flow luminosity
distance modulus of $\mu = 35.968$~mag\footnote{The host of \sn,
\host, is a likely member of the Hercules supercluster and may have a
large peculiar velocity.  A peculiar-velocity uncertainty of
1000~\kms\ corresponds to an uncertainty in the distance modulus of
0.012~mag.}.  Therefore, three out of the four distance fitters
provide values within 0.03~mag of the expected value if \host\ is in
the Hubble flow.  We can also measure the distance modulus with the
formula $\mu = V_{\rm peak} - M_{V, {\rm peak}} - A_{V, \rm{host}} -
A_{V, {\rm MW}}$.  Using the values measured and assumed by
\citetalias{Hicken09:de}, we find $\mu = 36.254$ and 35.826~mag for
$A_{V} = 0$ and 0.428~mag, respectively.  These values differ from the
Hubble-flow distance modulus by 0.286 and $-0.146$~mag, respectively,
corresponding to the offset from the WLR in Figure~\ref{f:phillips}.
Obviously, fitting all light curves of \sn\ provides a better estimate
of the distance to \sn\ than using only $V$-band data.

Although \sn\ appears to have a somewhat odd light curve, it is still
a relatively good standardizable candle.  \sn\ is both intrinsically
faint and red for its light-curve shape.  All light-curve fitters
correct for its red color by effectively brightening its apparent
magnitudes.  Since intrinsic color variation and dust extinction are
not orthogonal, deviations from the model in MLCS are regarded as dust
extinction (see \citealt{Conley07} for a discussion of how MLCS and
SALT treat color).  We discuss the implications of including \sn-like
objects in a cosmological analysis in Section~\ref{ss:cosmo}.


\section{Spectral Analysis}\label{s:spec}

In Figure~\ref{f:spec}, we show our collection of spectra for \sn.  In
this section, we examine the spectra in detail.

\subsection{Supernova Redshift}\label{ss:redshift}

\sn\ was located between \host\ and 2MASX~J15562803+2002482, two objects at
similar, but significantly different, recession velocities ($cz = 9628
\pm 55$ and $13,894 \pm 30$~\kms, respectively).  To determine the
most likely host galaxy, we cross-correlated all spectra obtained at
$t < 7$~days with a library of SN~Ia spectra using SNID
\citep{Blondin07}.  We determined two likely redshifts for each
spectrum: the redshift from the best-fitting template and the average
redshift from all SN~1991bg-like objects (since \sn\ has a very
similar spectrum to this class of objects; see
Section~\ref{ss:premax}). The best-fitting template spectrum was of
SN~1986G for all but one spectrum.  Performing these fits, we find
$cz_{\rm SN} = 9397 \pm 109$ and $8682 \pm 73$~\kms\ for the best-fit
template and SN~1991bg-like templates, respectively.  Considering the
velocity dispersion of the galaxy and the peculiarity of \sn, we
believe that these values are consistent with \sn\ being associated
with \host\ and inconsistent with \sn\ being associated with
2MASX~J15562803+2002482.  For the remainder of the paper, we assume
the redshift of \sn\ is that of \host.

\subsection{Pre-Maximum Spectrum}\label{ss:premax}

\begin{figure}
\begin{center}
\epsscale{1.15}
\rotatebox{0}{
\plotone{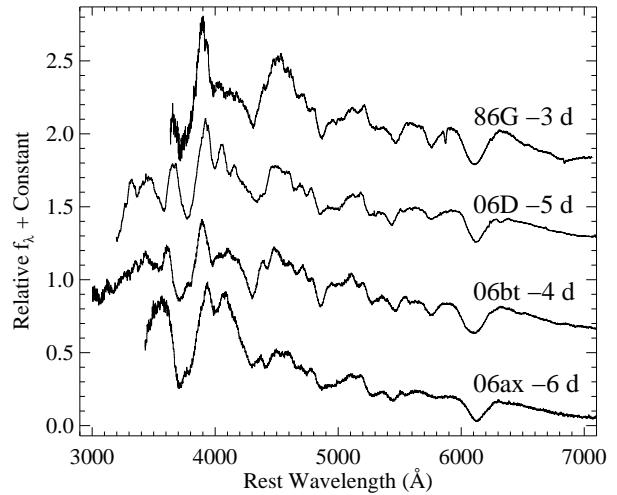}}
\caption{The pre-maximum-light spectrum of \sn\ compared to those of
SNe~1986G ($\Delta m_{15} (B) = 1.65$~mag), 2006D ($\Delta m_{15} (B)
approx 1.35$~mag) and 2006ax ($\Delta m_{15} (B) = 1.08$~mag) at a
similar phase.  The spectrum of SN~1986G has been dereddened by
$E(B-V)_{\rm MW} = 0.115$~mag and $E(B-V)_{\rm host} = 0.4$~mag.  The
spectrum of \sn\ shows many similarities to that of SN~1986G including
strong \ion{Si}{2}~$\lambda 5972$ and the presence of
\ion{Ti}{2}.}\label{f:cs}
\end{center}
\end{figure}

Our first spectrum was obtained 3 days before $B$-band maximum.  Using
this spectrum, \citet{Filippenko06:06bt} noted that \sn\ was somewhat
peculiar.  In Figure~\ref{f:cs}, we present this spectrum as well as
spectra of SNe~1986G ($\Delta m_{15} (B) = 1.65$~mag) and 2006ax
($\Delta m_{15} (B) = 1.08$~mag) at a similar phase.  Despite having
the same value of $\Delta m_{15} (B)$, SNe~2006ax and 2006bt exhibit
very different spectra at this epoch, with \sn\ having a much stronger
\ion{Si}{2} $\lambda 5972$ feature and a depression at \about4100~\AA\
corresponding to \ion{Ti}{2}.  These features are indicative of a
relatively cool photosphere and are hallmarks of the low-luminosity
SN~1991bg-like class of SNe~Ia \citep[e.g.,][]{Filippenko92:91bg,
Leibundgut93}.  Although there is some dispersion in the spectra of
objects with the same light-curve shape \citep{Benetti04, Matheson08},
differences of this magnitude have never before been seen within
normal SNe~Ia of the same light-curve shape.

The earliest spectrum of \sn\ is more similar to the spectrum of
SN~1986G, a low-luminosity SN~Ia and a member of the SN~1991bg-like
class of objects \citep{Phillips87}.  SN~1986G has stronger
\ion{Si}{2}~$\lambda 5972$ and \ion{Ti}{2} features, indicating that
\sn\ is slightly hotter than SN~1986G. (However, we do caution that 
the uncertain reddening correction of SN~1986G may cause the relative
strengths of its features to be incorrect.)

At this phase, the minimum of the \ion{Si}{2}~$\lambda 6355$
absorption feature is blueshifted by 12,500~\kms, which is the same as
SN~1986G and within the typical range of slower declining SNe~Ia
(although it is slightly larger than the velocity of SN~2006ax at that
phase).  The minimum of the \ion{Ca}{2} near-infrared triplet (seen in
Figures~\ref{f:spec} and \ref{f:synow}) is blueshifted by 15,900~\kms,
which is also normal.  There is no obvious high-velocity structure
associated with either feature.

Despite having a slow optical decline rate, \sn\ has a pre-maximum
spectrum most similar to those of fast decliners.  \sn\ has a
relatively cool photosphere, which may explain the lack of a shoulder
in $r$ and a prominent second maximum in $i$ \citep{Kasen06}, but is
contradictory given the slow decline \citep{Kasen07:wlr}.

There is no indication of Na~D absorption at either zero velocity
(consistent with the low value of $E(B-V)$ measured by
\citealt{Schlegel98}) or the host-galaxy recession velocity in any of
our spectra, indicating a small or negligible extinction.

\subsection{Late-Time Spectrum}

\begin{figure}
\begin{center}
\epsscale{1.15}
\rotatebox{0}{
\plotone{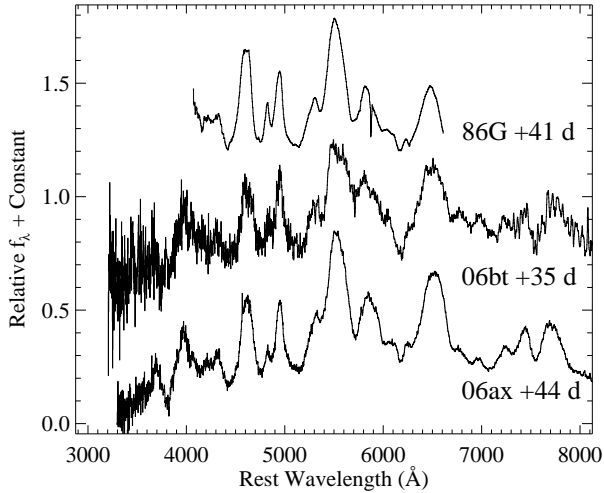}}
\caption{The $t = +35$~day spectrum of \sn\ compared to those of
SNe~1986G ($\Delta m_{15} (B) = 1.65$~mag) and 2006ax ($\Delta m_{15}
(B) = 1.08$~mag) at a similar phase.  The spectrum of SN~1986G has
been dereddened by $E(B-V)_{\rm MW} = 0.115$~mag and $E(B-V)_{\rm
host} = 0.4$~mag.  At this phase, all three SNe have similar
spectra.}\label{f:cs2}
\end{center}
\end{figure}

In Figure~\ref{f:cs2}, we present our $t = 35$~day spectrum compared
to spectra of SNe~1986G and 2006ax at a similar phase.  The
differences between SN spectra become smaller with time, and by $t =
35$~days, the spectrum of \sn\ is very similar to spectra of both
SNe~1986G and 2006ax at this time.

\subsection{Quantitative Measurements}

There have been several attempts to classify SNe~Ia into distinct
groups based on their spectra \citep[e.g.,][]{Nugent95, Benetti05,
Branch06}.  Most methods rely (at least in one dimension) on the
strength of \ion{Si}{2}~$\lambda 5972$ relative to that of
\ion{Si}{2}~$\lambda 6355$.  This is easily measured by the parameter
$\mathcal{R}$(Si) \citep{Nugent95}, but \citet{Branch06} advocates
measuring the equivalent widths of these lines.  In addition to
$\mathcal{R}$(Si), \citet{Benetti05} divided SNe~Ia based on a
combination of light-curve shape and the velocity gradient of the
\ion{Si}{2}~$\lambda 6355$ feature, $\dot{v}$.  \citet{Branch06}
defines four subclasses, namely shallow silicon (SS), core normal
(CN), broad line (BL), and cool (CL), while \citet{Benetti05} defines
three subclasses called low-velocity gradient (LVG), high-velocity
gradient (HVG), and faint (FAINT).  The classification systems have
much overlap, with SS and CN objects mostly being in the LVG, BL
objects mostly being in the HVG, and CL mapping directly to FAINT.

\begin{figure}
\begin{center}
\epsscale{1.15}
\rotatebox{0}{
\plotone{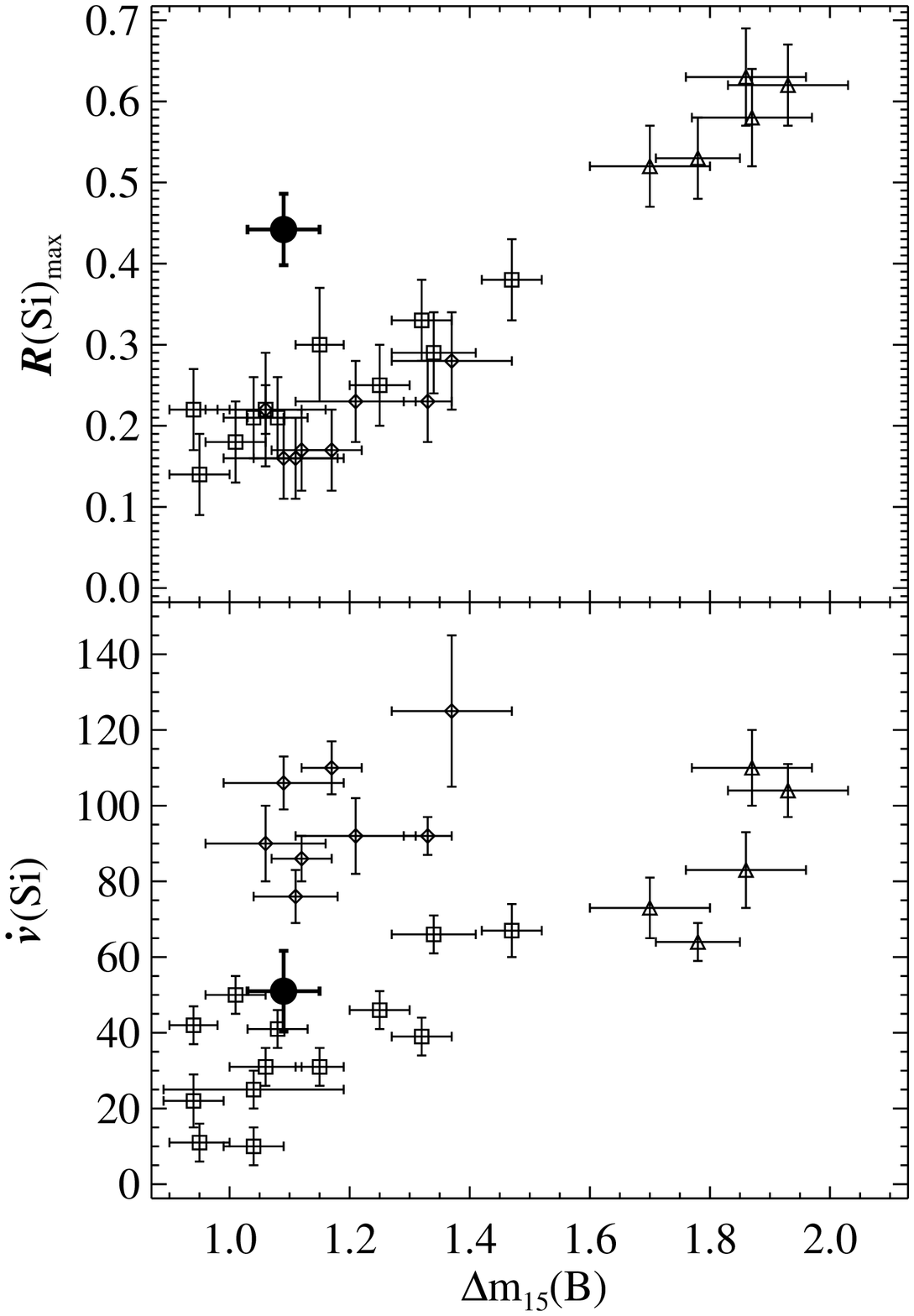}}
\caption{Silicon ratio at maximum brightness ({\it top};
$\mathcal{R}$(Si)$_{\rm max}$) and velocity gradient ({\it bottom};
$\dot{v}$(Si)) vs.\ $\Delta m_{15} (B)$ for a sample of SNe~Ia (taken
from \citealt{Benetti05}) and \sn.  The LVG, HVG, and FAINT groups (as
defined by \citealt{Benetti05}) are represented by squares, diamonds,
and triangles, respectively.  \sn\ is shown by the filled circle.
Based on the velocity gradient, \sn\ is a member of the LVG group of
SNe~Ia despite having a spectrum similar to those of the FAINT group.
\sn\ is a clear outlier in the relationship between the light-curve
shape and the ratio of the silicon features.}\label{f:si}
\end{center}
\end{figure}

At maximum light, \sn\ has $W(5750) = 29$~\AA, $W(6100) = 120$~\AA,
$\mathcal{R}{\rm (Si)} = 0.44$, and $\dot{v} = 51$\kms~day$^{-1}$.
These values place \sn\ within the LVG group (although toward the high
end of the range of $\dot{v}$; see Figure~\ref{f:si}) and between the
CL and BL groups, with its closest neighbors (using the updated values
of \citealt{Branch09}) being the BL SN~1999cc ($W(5750) = 26$~\AA;
$W(6100) = 121$~\AA) followed by CL SN~1989B ($W(5750) = 25$~\AA;
$W(6100) = 126$~\AA).  Despite being closest to a BL object in the
equivalent width space, the shape of the \ion{Si}{2} absorption and
the borderline values should place \sn\ in the CL group.

The true peculiar nature of \sn\ is apparent when comparing a
temperature-dependent spectral parameter such as $\mathcal{R}$(Si)
with a light-curve parameter such as $\Delta m_{15} (B)$.  In
Figure~\ref{f:si}, we show these values for the sample of
\citet{Benetti05} (open symbols) and for \sn\ (filled circle).  In
this parameter space, \sn\ is a large outlier, indicating that it has
a relatively cool photosphere and a slow decline rate.

\subsection{Possible Detection of Carbon}\label{ss:carbon}

\begin{figure}
\begin{center}
\epsscale{1.15}
\rotatebox{0}{
\plotone{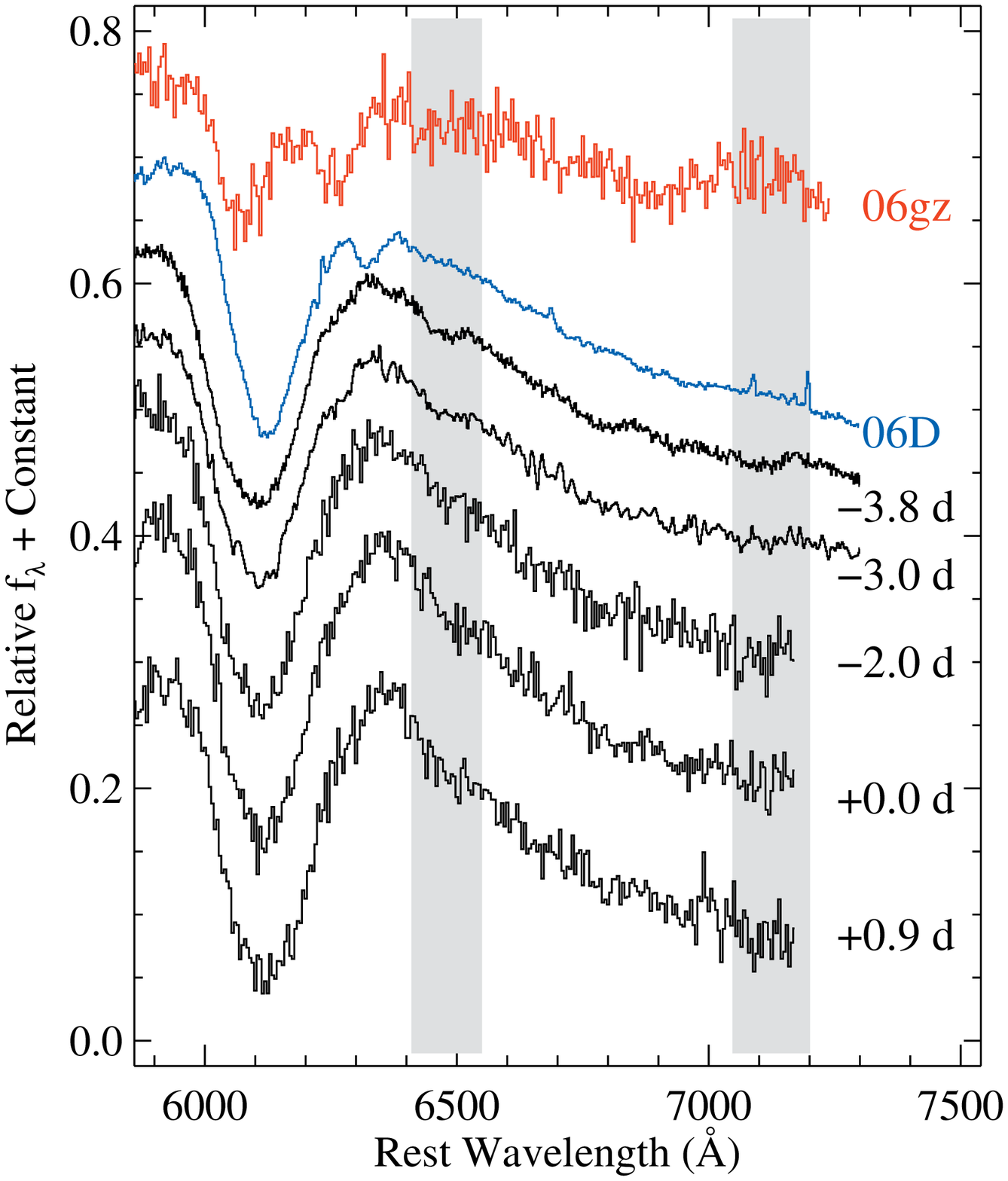}}
\caption{Spectra of \sn\ (black curves) near the suspected
\ion{C}{2}~$\lambda\lambda 6580$, 7234 features.  Spectra of SNe~2006D
($t = -5$~days; \citealt{Thomas07}) and 2006gz ($t = -14$~days;
\citealt{Hicken07}) are also shown for comparison (blue and red
curves, respectively).  Wavelengths near the potential \ion{C}{2}
features (corresponding to a blueshifted velocity of \about
1350--7850~\kms) have been shaded grey.  The \ion{C}{2} features of
SNe~2006D and 2006gz have blueshifted velocities of 12,000 and
15,500~\kms, respectively.  [{\it See the electronic edition of the
Journal for a color version of this figure.}]}\label{f:carbon}
\end{center}
\end{figure}

In our earliest spectra, there is a small feature redward of
\ion{Si}{2}~$\lambda 6355$ (see Figure~\ref{f:carbon}).  In our first
spectrum (when the feature is strongest and the signal-to-noise ratio
of the spectrum is highest), its minimum is at 6465~\AA.  This feature
is present in our second and third spectra, although with less
significance.  These three spectra come from different instruments and
telescopes, indicating that it is not an artifact of data reduction.
The feature has mostly disappeared in our fourth spectrum ($t =
0.0$~days).

An intriguing possibility is that this feature is from
\ion{C}{2}~$\lambda 6580$ absorption.  If this is the case, the
minimum of the absorption would be blueshifted by 5200~\kms, which is
much lower than that of other features and may indicate a
misidentification.  This feature has been detected in a handful of
objects \citep[e.g.,][]{Branch03} with the best cases being SNe~2006D
\citep{Thomas07}, 2006gz \citep{Hicken07}, and 2009dc
\citep{Yamanaka09}, but all detections so far have been at much larger
velocities and typically at much earlier phases.  For instance, the
minimum of the \ion{C}{2}~$\lambda 6580$ feature in SN~2006gz was
blueshifted by \about 15,500~\kms\ at $t = -14$~days (which was a
larger velocity than the \ion{Si}{2} feature at that phase) and had
completely disappeared by $t = -12$~days.

\ion{C}{2} has four relatively strong features in the optical:
$\lambda\lambda 4267$, 4745, 6580, and 7234.  Unfortunately,
\ion{C}{2}~$\lambda 4267$ would be on the edge of a strong \ion{Mg}{2}
feature and \ion{C}{2}~$\lambda 4745$ is on top of a \ion{Ti}{2}
feature (although it may still be detectable with SYNOW; see
Section~\ref{ss:synow}).  Only the two reddest lines are in relatively
uncontaminated regions of the spectrum.  As seen in
Figure~\ref{f:carbon}, there is a very shallow feature in our earliest
spectrum at the wavelength that corresponds to \ion{C}{2}~$\lambda
7234$ blueshifted by \about 5000~\kms.

Observations have shown that high-velocity features are common in
SNe~Ia \citep[e.g.,]{Mazzali05}.  The data can be explained by
circumstellar interaction, dense blobs of material ``detached'' from
the rest of the ejecta, or a differing covering fraction of the ejecta
\citep{Gerardy04, Mazzali05, Tanaka06}.  If there is a single carbon
blob ejected at an angle away from our line of sight, the absorption
velocity seen in the spectra would correspond to the component of the
velocity vector along our line of sight.  For an angle of 45\arcdeg\
(65\arcdeg) relative to our line of sight, we infer a true velocity of
7450~\kms\ (12,300~\kms) in such a scenario.  This potential asymmetry
may help explain other aspects of this SN (see
Section~\ref{ss:models}).  Alternatively, there may be a significant
amount of carbon-rich material close to the SN.  For this scenario,
viewing-angle effects necessitate that our line of sight be in the
plane of the circumstellar accretion disk.

Besides the possible detection of carbon lines, \sn\ shares other
characteristics with SNe~2006D and 2006gz.  SN~2006D had a relatively
narrow light curve similar to that of SN~1986G, but not as narrow as
that of SN~1991bg \citep{Thomas07}.  The $t = -5$~day spectrum from
\citet{Thomas07} is reproduced in Figure~\ref{f:cs}.  The early-time
spectra of SNe~2006D and 2006bt are relatively similar, both having
relatively large $\mathcal{R}$(Si).  \sn\ has slightly larger
velocities for the Si and Ca features and stronger \ion{Ti}{2}
features than SN~2006D.

SN~2006gz had very broad light curves ($\Delta m_{15} (B) = 0.69 \pm
0.04$), no shoulder in $r$, and no second maximum in $i$
\citep{Hicken07}.  The colors were also very red, but
\citet{Hicken07} attributed this to a large amount of host-galaxy
extinction ($E(B-V)_{\rm host} = 0.18$~mag).  SN~2006gz had spectra
indicative of a very hot photosphere unlike \sn.

\subsection{SYNOW Model Fits}\label{ss:synow}

To investigate the details of our SN spectra, we use the supernova
spectrum-synthesis code SYNOW \citep{Fisher97}.  Although SYNOW has a
simple, parametric approach to creating synthetic spectra, it is a
powerful tool to aid line identifications which in turn provide
insights into the spectral formation of the objects. To generate a
synthetic spectrum, one inputs a blackbody temperature ($T_{\rm BB}$),
a photospheric velocity ($v_{\rm ph}$), and for each involved ion, an
optical depth at a reference line, an excitation temperature ($T_{\rm
exc}$), the maximum velocity of the opacity distribution ($v_{\rm
max}$), and a velocity scale ($v_{e}$). This last variable assumes
that the optical depth declines exponentially for velocities above
$v_{\rm ph}$ with an $e$-folding scale of $v_{e}$.  The strengths of
the lines for each ion are determined by oscillator strengths and the
approximation of a Boltzmann distribution of the lower-level
populations with a temperature of $T_{\rm exc}$.

\begin{figure}
\begin{center}
\epsscale{1.15}
\rotatebox{0}{
\plotone{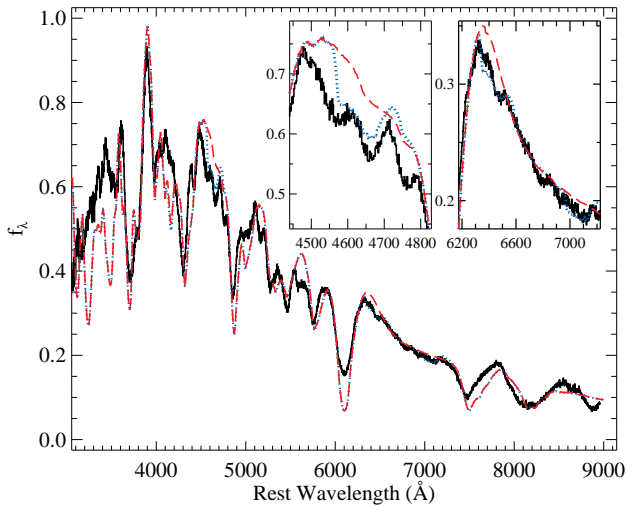}}
\caption{Our $-3.8$~day optical spectrum of \sn\ (black) and our
best-fit SYNOW synthetic spectrum with and without \ion{C}{2} (blue
dotted and red dashed lines, respectively).  The insets show the
regions near \ion{C}{2} $\lambda\lambda 4745$, 6580, 7234.  The fit is
improved slightly by including \ion{C}{2}.  [{\it See the electronic
edition of the Journal for a color version of this
figure.}]}\label{f:synow}
\end{center}
\end{figure}

In Figure~\ref{f:synow}, we present our $-3.8$~day spectrum of \sn\
with a synthetic spectrum generated from SYNOW.  This fit is very
similar to the fit performed for SN~1986G by \citet{Branch06} and the
optical depths for the fit are presented in Table~\ref{t:synow}.  The
main differences are that we include \ion{O}{1} (it was excluded from
the reference fit because of the short wavelength range of the
SN~1986G spectrum), the excitation temperature is 8000~K instead of
7000~K, all ions other than \ion{Mg}{1} and \ion{Mg}{2} have lower
optical depths, we do not include high-velocity \ion{Ca}{2} and
\ion{Fe}{2}, and we do not impose a maximum velocity for \ion{Si}{2}.
The model includes low-ionization ions and excludes high-ionization
ions, confirming the low-temperature nature of \sn.  No ions at high
velocity are included, further limiting the possibility of
high-velocity ejecta.  There are some differences between the \sn\ and
SYNOW spectra.  It is possible that these differences could be reduced
by examining more of the parameter space of SYNOW input parameters.
These differences do not affect our conclusions below.

\begin{deluxetable*}{lcccccccccccc}
\tabletypesize{\scriptsize}
\tablewidth{0pt}
\tablecaption{SYNOW Optical Depths\label{t:synow}}
\tablehead{
\colhead{} &
\multicolumn{12}{c}{Value} \\
\colhead{Parameter} &
\colhead{\ion{C}{2}} &
\colhead{\ion{O}{1}} &
\colhead{\ion{Mg}{1}} &
\colhead{\ion{Mg}{2}} &
\colhead{\ion{Si}{2}} &
\colhead{\ion{S}{2}} &
\colhead{\ion{Ca}{1}} &
\colhead{\ion{Ca}{2}} &
\colhead{\ion{Sc}{2}} &
\colhead{\ion{Ti}{2}} &
\colhead{\ion{Co}{2}} &
\colhead{\ion{Ni}{2}}}

\startdata

$\tau$        & 2 & 5 & 3 & 8 & 50 & 1 & 3 & 35 & 0.5 & 0.8 & 0.5 & 0.4 \\

\enddata

\end{deluxetable*}

A slightly higher excitation temperature and lower optical depths in
\sn\ compared to SN~1986G is consistent with our relatively crude
assessment of the spectrum in Section~\ref{ss:premax} where we
considered \sn\ to be slightly hotter (and more ``normal'') than
SN~1986G.  The model for SN~1986G has $v_{\rm max} = 15,000$~\kms,
while the model for \sn\ imposes no limit.  Where nuclear burning may
only extend to $v = 15,000$~\kms\ in SN~1986G, it appears that it
extends to all velocities in \sn.

Since we may have identified \ion{C}{2} in the pre-maximum spectra of
\sn\ (see Section~\ref{ss:carbon}), we created models with and without
\ion{C}{2}.  In an attempt to match the possible low-velocity
features, we included \ion{C}{2} with $T_{\rm exc} = 12,000$~K,
$v_{\rm max} = 15,000$~\kms, and $v_{e} = 200$~\kms.  The inclusion of
\ion{C}{2} with these parameters improves the model fit.  In addition
to better fits near \ion{C}{2}~$\lambda\lambda 6580$, 7234, the model
is significantly better near \ion{C}{2}~$\lambda 4745$.  This shows
that it is possible to detect lines in confused regions such as near
4600~\AA.  \citep{Branch06} fit this region with high-velocity
\ion{Fe}{2}, and again, we caution that the identification of
\ion{C}{2} is not definitive.


\section{Supernova Environment}\label{s:enviro}

\begin{figure*}[ht]
\begin{center}
\epsscale{0.6}
\rotatebox{90}{
\plotone{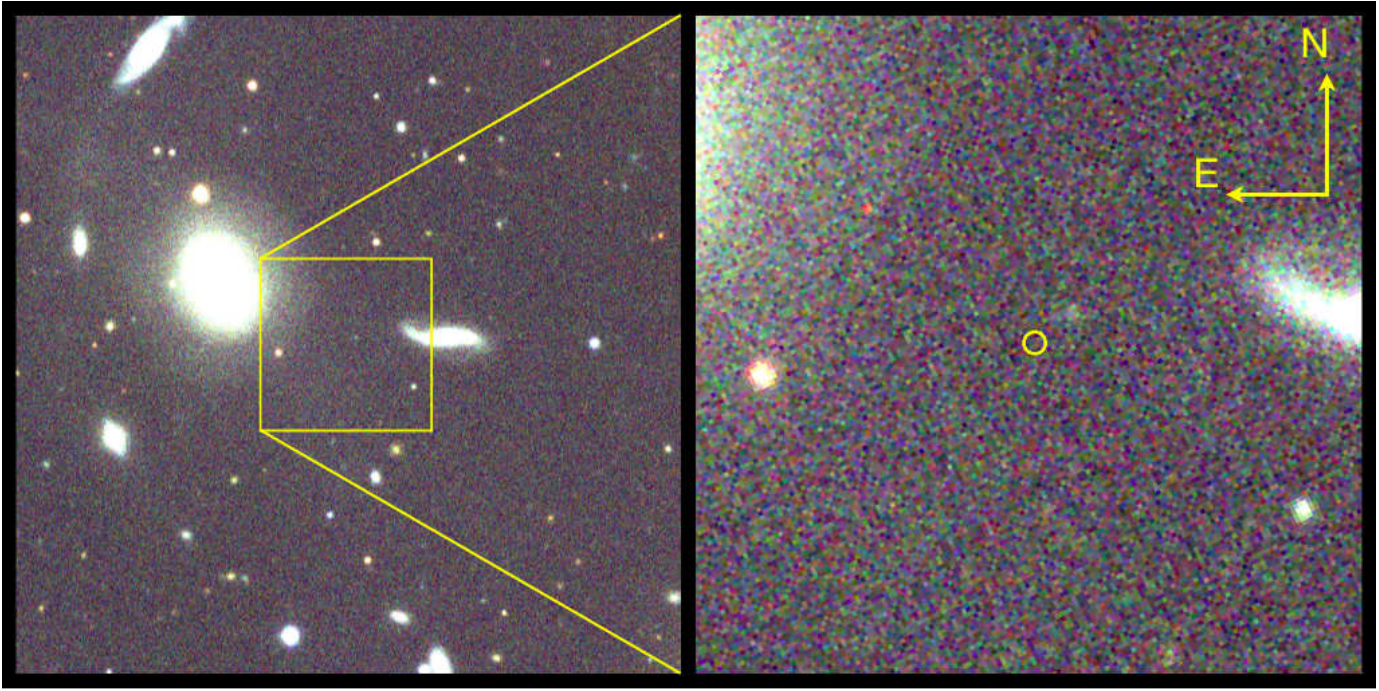}}
\caption{({\it left}) SDSS \gri false-color image of the position of
\sn\ and its surrounding environment.  The image is 4\arcmin $\times
4$\arcmin\ in extent.  The yellow box is 1\arcmin\ on a side and
centered on the SN position.  The SN has a projected galactocentric
distance of 33.7~kpc relative to its host galaxy, \host, the large
galaxy on the northeast edge of the yellow box.  The nearby galaxy to
the west has a recession velocity 4250~\kms\ greater than that of the
nominal host galaxy and does not appear to be interacting with the
host galaxy.  ({\it right}) Zoomed-in image of the SN position
corresponding to the yellow box on the left (1\arcmin $\times
1$\arcmin).  The SN position is marked by a circle of 1\arcsec\
radius.  There is a faint source 4.3\arcsec\ (corresponding to 2.9~kpc
at the redshift of \host) to the northwest that is likely to be a
background galaxy (see text for details).  [{\it See the electronic
edition of the Journal for a color version of this figure.}]
}\label{f:finder}
\end{center}
\end{figure*}

\sn\ occurred 33.7~kpc ($7.8 R_{\rm eff}$, where $R_{\rm eff} =
6.4\arcsec$ is defined by the Petrosian radius where 50\% of the flux
is contained within a circle of radius $R_{\rm eff}$) from the nucleus
of its host galaxy, \host, an S0/a galaxy (see Figure~\ref{f:finder}).
\sn\ lies between its host and a nearby galaxy,
2MASX~J15562803+2002482 (\sn\ is 23.7\arcsec\ east of the nucleus of
2MASX~J15562803+2002482; although \sn\ lies between these two
galaxies, we are confident that \host\ is the host galaxy; see
Section~\ref{ss:redshift}).  This galaxy has $z = 0.0463$,
corresponding to a recession velocity that is 4250~\kms\ greater than
that of \host\ \citep{Abazajian09}.  Examining SDSS images of the SN
environment \citep{Abazajian09}, there does not appear to be any tidal
streams between the two galaxies.  Although both objects are likely
members of the Hercules supercluster, it is unlikely that they have
interacted.  There is another faint source 4.3\arcsec\ to the
northwest (corresponding to 2.9~kpc at the redshift of \host) that has
colors consistent with a background galaxy (with SDSS photo-$z$ of
$0.36 \pm 0.17$; \citealt{Csabai03}) or a globular cluster.  If it
were a globular cluster at the redshift of \host, it would have $M_{g}
= -12.1$~mag, which is \about 2~mag brighter than G1 in M31.  It is
therefore more likely to be a background galaxy.

\begin{figure}
\begin{center}
\epsscale{1.15}
\rotatebox{0}{
\plotone{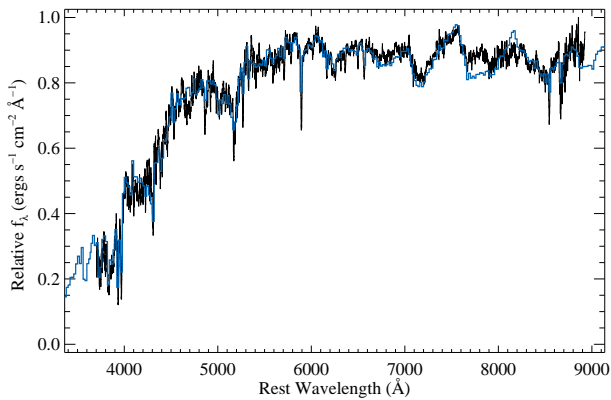}}
\caption{Spectrum of \host\ (black) and the best-fit SSP model ($t =
14$~Gyr; $Z = 0.012 = 0.6 Z_{\sun}$; blue).}\label{f:galspec}
\end{center}
\end{figure}

A spectrum of the nucleus of \host\ from SDSS \citep{Abazajian09}
shows no emission lines, suggesting little or no recent star
formation.  Using the single-stellar populations (SSP) models of
\citet{Jimenez04}, the spectrum of \host\ is best fit by a population
with a power-law initial mass function, an age of 14~Gyr, and a
metallicity of 0.6 solar (see Figure~\ref{f:galspec}).  It is worth
noting that \citet{Jimenez04} has no model with an age $>14$~Gyr.
Although such ages are limited by the age of the Universe, if the
model ages are systematically larger than true ages, we may not be
probing the full range of realistic models.  Despite the excellent fit
with an old SSP model, we cannot rule out additional (low-level) star
formation over the past several Gyr.

Given the position of the SN, it is likely that the progenitor of \sn\
was a member of the halo population.  If this were the case, then the
progenitor would likely have formed before the majority of the stars
in the bulge of the galaxy and with a lower metallicity.  In this
scenario, it is quite likely that the progenitor was of low
metallicity and had an age of $> 10$~Gyr.

Alternatively, the progenitor may have formed in a galactic disk or
the bulge and was ejected outward from an interaction with a
supermassive black hole at the nucleus of the galaxy.  To reach
33.7~kpc, the progenitor would have to travel at 3300, 330, 33, and
6.6~\kms\ for 10~Myr, 100~Myr, 1~Gyr, and 5~Gyr, respectively.  From
its position alone, it is very unlikely that the progenitor of \sn\
was a massive star ($M > 10 {\rm M}_{\sun}$) that formed near the
galactic center and was ejected into the halo.  It is possible that
the progenitor was a low-mass star ejected in this manner, which would
probably make it younger and have higher metallicity than other stars
in the halo.  Regardless, it is very likely that the progenitor of
\sn\ was a white dwarf and \sn\ was not the result of some exotic
core-collapse event.


\section{Discussion}\label{s:disc}

\subsection{Models}\label{ss:models}

It has long been believed that the vast majority of SN~Ia light curves
can be described by a single parameter which directly correlates with
the amount of $^{56}$Ni produced in the explosion (with more $^{56}$Ni
corresponding to more luminous events having broader light curves).
\citet{Kasen07:wlr} explain this further by noting that objects with
larger $M_{\rm Ni}$ are hotter, making the recombination from
\ion{Fe}{3} to \ion{Fe}{2} occur at a later time, and thus have a
broader $B$-band light curve.

\citet{Kasen06} found that the depth of the trough, the height of the
second maximum, and the delay between the first and second maxima in
the $I$ band all increase with $^{56}$Ni mass.  This finding would
imply that \sn\ had a small $^{56}$Ni mass (model $I$-band light
curves with $0.2 \le M_{\rm Ni} \le 0.3 {\rm M}_{\sun}$ show a similar
behavior to the $i$-band light curve of \sn), which is in direct
conflict with the decline rate of all bands.

The prominence of the second maximum is more generally related to the
mass of all Fe-group elements in the ejecta, not just $^{56}$Ni and
the isotopes into which it decays \citep{Kasen06}.  The second maximum
can be reduced by decreasing the progenitor metallicity, which
decreases the amount of stable Fe-group elements (``stable Fe'')
produced in the explosion \citep{Timmes03}; however, the effect is at
most \about 10\%.  Since stable Fe is produced via electron capture,
it is possible that the explosion simply did not have the physical
conditions necessary to form a significant amount of stable Fe through
electron capture.  Finally, one can smear out the second maximum in
$I$ by mixing the stable Fe and $^{56}$Ni throughout the ejecta
\citep{Kasen06}.  This will also cause the maximum of the $I$ band to
occur later, which may be the case for \sn.

\sn\ may have been a very asymmetric explosion.  Although we see no
indication of high-velocity ejecta in any of the spectral features,
there is the possibility that a carbon blob was ejected \about
65\arcdeg\ from our line of sight.  SN~1999by, a low-luminosity event,
showed a very polarized spectrum corresponding to an explosion with an
asphericity of 20\% viewed along the equator \citep{Howell01:99by}.
Asymmetric models show that peak luminosity depends on viewing angle;
however, these models do not exhibit significant differences in
ionization with viewing angle \citep{Kasen07:asym, Sim07}.

It is also possible that \sn\ rose very quickly to maximum, causing a
high luminosity with a smaller amount of $^{56}$Ni \citep{Arnett82}.
There are poor constraints on an explosion date from pre-explosion
images \citep{Lee06}.  Our photometry starts only a few days before
$B$ maximum, but it does not appear to be faster than other objects
with its decline rate.  Although a fast rise is still a possibility,
it appears to be an unlikely solution.

\subsection{Implications for Cosmology}\label{ss:cosmo}

Given its relatively normal luminosity, \sn-like objects would still
be detected in high-redshift SN surveys, unlike other peculiar objects
such as SNe~2002cx \citep{Li03:02cx} and 2008ha \citep{Foley09:08ha,
Valenti09}.  Furthermore, given its $B-V$ color curve, a \sn-like
object with some host-galaxy extinction will simply look like a normal
object with that extinction plus an additional $A_{V} \approx 0.5$~mag
of extinction.  At some point too much extinction would lead to the
exclusion of these objects from a SN sample; most cosmological
analyses avoid SNe~Ia having extremely red colors.

Additionally, at high redshift, the rest-frame $r$ and $i$ band will
be redshifted out of the optical, requiring observers to rely on
rest-frame $B-V$ alone to determine the quality of the light-curve
fits and measure the extinction.  At low redshift, one could search
for objects with large $\Delta m_{15} (B)$ but no secondary maximum in
$i$ and exclude any such objects; however, at high redshift this is
not feasible without observing in the near infrared ({\it JWST} and,
depending on the ultimately chosen design, JDEM might be able to
measure the rest-frame $i$ band out to $z \approx 1$).  For an object
at $z \gtrsim 0.3$ (where $i$ is redshifted out of the optical), one
would have to rely on a spectrum (in combination with a light curve)
to identify \sn-like objects.  However, with the \ion{Si}{2} lines
redshifting out of the optical window at $z \approx 0.5$ and the
generally low signal-to-noise ratio spectra of high-redshift SNe~Ia,
it is difficult to detect spectroscopically peculiar objects from
high-redshift surveys \citep{Foley09:year4}.

If \sn\ does come from a very old progenitor and the age of the
progenitor is the main reason for its peculiarity, then \sn-like
objects may not exist at high redshift.  At $z = 0.5$ and 1, the
Universe had an age of \about 8.6 and 5.9~Gyr, respectively.  If the
delay time between formation and explosion is longer than those times,
\sn-like objects will not exist at those redshifts. \citet{Foley09:year4} 
found no SN~1991bg-like objects in a sample of 118 high-redshift
SNe~Ia, indicating that \sn-like objects are probably not a large
fraction of the early-universe SN~Ia population.

We have developed a Monte Carlo simulation to assess the impact of
contamination of a population of \sn-like objects in a SN~Ia
cosmological sample.  For normal SNe~Ia, we generate 1000 light curves
over a range of light-curve shape parameters from the MLCS2k2
``early'' templates \citep{Jha07} over $0 < z < 0.8$ (500 for $z <
0.1$ and 500 for $z > 0.1$), with distance modulus, $\mu_{\rm true}$,
calculated in a nominal flat $\Lambda$CDM ($\Omega_{M} = 0.3$, $h =
0.65$) cosmology.  We parameterize the input extinction by a $R_{V} =
3.1$ \citet{ODonnell94} law, with $A_{V}$ drawn from the {\it glos}
distribution \citep[e.g.,][]{Hatano98}.  The light-curve shape
parameter $\Delta$ is drawn randomly between $-0.3$ and 1.3.  The
sampling and signal-to-noise ratio is matched to the expected values
for the Pan-STARRS Medium Deep Survey.

Since \sn\ has very different light curves from those of other SNe~Ia,
the MLCS2k2 template light curves had to be modified to create
simulated light curves of \sn-like objects.  Specifically, the
\sn-like light curves must be intrinsically underluminous and red
compared to a nominal SN~Ia with the same light-curve shape.  To
determine the appropriate modifications, each individual light curve
of \sn\ was fit separately.  The \ubvrr\ light curves all had best
fits of $\Delta \approx -0.15$, while the $i$-band light curve was
best fit with $\Delta = -0.38$.  The MLCS2k2 model light curves were
shifted so that the peak magnitudes were the same as those of \sn\ in
each band.  We created 100 \sn-like light curves for $0 < z < 0.8$ (50
for $z < 0.1$ and 50 for $z > 0.1$).

All light curves were fit using MLCS2k2 \citep{Jha07}, which provided
estimates of $A_{V}$, $\Delta$, and $\mu$.  The fitting was performed
with both the {\it glos} and {\it flatnegav} (a flat prior which
includes negative values of $A_{V}$) priors.  Fits of all light
curves, including those for \sn-like objects, had a reduced $\chi^{2}
\leq 2$, indicating that all would nominally be included in a
cosmological analysis.

\begin{figure}
\begin{center}
\epsscale{1.15}
\rotatebox{0}{
\plotone{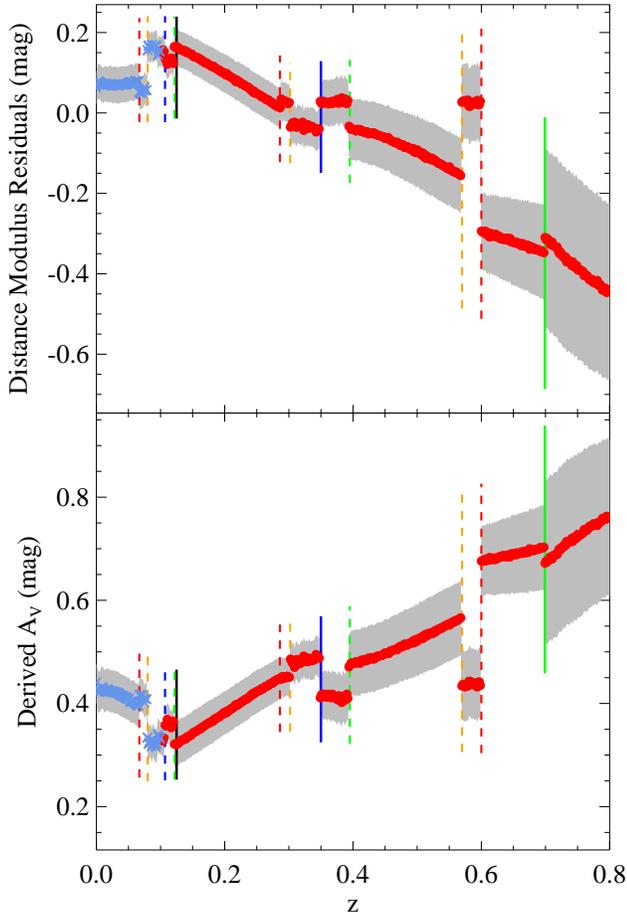}}
\caption{Distance modulus residuals ({\it top}; $\mu_{\rm MLCS} -
\mu_{\rm true}$) and derived $A_{V}$ ({\it bottom}) from MLCS2k2 as a
function of redshift for the noise-free model light curves for \sn.
The blue and red points represent objects with redshifts less than and
greater than $z = 0.1$, respectively.  The grey regions are the model
uncertainties.  There are several discontinuities at particular
redshifts corresponding to either an observed-frame filter no longer
matching a rest-frame filter (e.g., the observed $V$ dropping out at
$z = 0.699$; solid lines) or a transition from an observed-frame
filter being K-corrected to a particular filter changing to a
different one (e.g., observed $i$ being K-corrected to rest-frame $V$
switching to rest-frame $B$ at $z = 0.570$; dashed lines).  The black,
blue, green, red, and orange lines represent the rest-frame \ubvri\
filters undergoing transitions, respectively.  [{\it See the
electronic edition of the Journal for a color version of this
figure.}]\label{f:resid}}
\end{center}
\end{figure}

To demonstrate how MLCS2k2 interprets our \sn-like model light curves
as a function of redshift, we produced noise-free light curves
spanning the redshift interval $0 < z < 0.8$.  The light curves were
fit by MLCS2k2 with a {\it flatnegav} prior (in this signal-to-noise
ratio regime, the prior is inconsequential and the uncertainty is
dominated by the MLCS model uncertainty).  The resulting distance
modulus residuals and derived $A_{V}$ are presented in
Figure~\ref{f:resid}.  The derived distance modulus residuals are
close to zero near $z = 0$, but increase to \about 0.1~mag at $z =
0.1$.  After that, the residuals generally linearly decrease with
increasing redshift, deviating by \about 0.5~mag at $z = 0.8$.  At
redshifts where either the mapping of observed filters to rest-frame
filters during the K-correction process changes or when a rest-frame
filter is no longer used in K-corrections (the filter ``drops out''),
there are large discontinuities in the residuals.  These
discontinuities are partly the result of our model for \sn\ and the
normal SN~Ia template model and partly the result of having the
observer-frame light curves derived from a single rest-frame filter.

The derived value for $A_{V}$ has a very similar (but inverted)
function with redshift.  Since our model assumes $A_{V} = 0$~mag at
all redshifts, the derived value corresponds to the difference between
the true distance modulus and the derived distance modulus, modulo the
difference in peak absolute magnitude for \sn\ and the canonical SN~Ia
with the same $\Delta$.  Although the distance modulus residuals
appear to be quite large, the distance modulus is compared to the
input cosmological model.  When fitting an observed sample, one can
only look for outliers from the best fit, and the inclusion of
\sn-like objects will influence such a fit.  Furthermore, a deviation
of a few tenths of a magnitude from the residual of the best-fit
cosmology is expected for several objects in a high-redshift SN
sample.

To examine the impact of \sn-like objects on cosmological parameter
estimation, we defined 7 distinct samples.  The first sample had no
contamination from \sn-like objects.  The remaining samples had 1\%,
5\%, or 10\% contamination at all redshifts, or only for $z < 0.1$
(consistent with a very long delay time for such objects).  The
distance moduli found using MLCS2k2 were used to constrain the
cosmological parameters $\Omega_{M}$ and $w$ with the assumption of
flatness.  We also combined the SN data with the prior from the baryon
acoustic oscillations \citep[BAO;][]{Eisenstein05}, further
constraining the cosmological parameters.  Since we fit the light
curves using two separate extinction priors, we have a total of 14
different estimates of $\Omega_{M}$ and $w$ with and without BAO
constraints.  The parameters are listed in Table~\ref{t:sim}.

\begin{deluxetable*}{lllll}
\tabletypesize{\scriptsize}
\tablewidth{0pt}
\tablecaption{Cosmological Parameter Recovery From Simulations\label{t:sim}}
\tablehead{
\colhead{} &
\colhead{$\Omega_{M}$} &
\colhead{$\Omega_{M}$} &
\colhead{$w$} &
\colhead{$w$} \\
\colhead{Sample} &
\colhead{(SNe Only)} &
\colhead{(SNe + BAO)} &
\colhead{(SNe Only)} &
\colhead{(SNe + BAO)}}

\startdata

\cutinhead{{\it flatnegav} Prior}

No   contamination         & $0.318 \err{0.097}{0.203}$ & $0.292 \err{0.028}{0.018}$ & $-1.079 \err{0.346}{0.479}$ & $-0.893 \err{0.069}{0.072}$ \\
1\%  contamination         & $0.322 \err{0.097}{0.206}$ & $0.293 \err{0.028}{0.018}$ & $-1.076 \err{0.347}{0.479}$ & $-0.883 \err{0.069}{0.071}$ \\
5\%  contamination         & $0.314 \err{0.106}{0.202}$ & $0.296 \err{0.028}{0.019}$ & $-1.013 \err{0.322}{0.485}$ & $-0.857 \err{0.067}{0.069}$ \\
10\% contamination         & $0.336 \err{0.105}{0.217}$ & $0.302 \err{0.029}{0.019}$ & $-1.014 \err{0.334}{0.487}$ & $-0.813 \err{0.065}{0.066}$ \\
1\%  contamination low-$z$ & $0.308 \err{0.103}{0.197}$ & $0.293 \err{0.028}{0.018}$ & $-1.037 \err{0.326}{0.482}$ & $-0.886 \err{0.069}{0.071}$ \\
5\%  contamination low-$z$ & $0.276 \err{0.121}{0.179}$ & $0.294 \err{0.029}{0.019}$ & $-0.921 \err{0.266}{0.471}$ & $-0.864 \err{0.067}{0.068}$ \\
10\% contamination low-$z$ & $0.238 \err{0.139}{0.156}$ & $0.296 \err{0.028}{0.019}$ & $-0.795 \err{0.200}{0.434}$ & $-0.832 \err{0.064}{0.065}$ \\

\cutinhead{{\it glos} Prior}

No   contamination         & $0.311 \err{0.082}{0.182}$ & $0.289 \err{0.027}{0.018}$ & $-1.094 \err{0.324}{0.415}$ & $-0.927 \err{0.067}{0.070}$ \\
1\%  contamination         & $0.278 \err{0.121}{0.180}$ & $0.295 \err{0.029}{0.019}$ & $-0.921 \err{0.268}{0.476}$ & $-0.861 \err{0.067}{0.069}$ \\
5\%  contamination         & $0.206 \err{0.155}{0.136}$ & $0.302 \err{0.029}{0.019}$ & $-0.678 \err{0.144}{0.368}$ & $-0.774 \err{0.060}{0.060}$ \\
10\% contamination         & $0.197 \err{0.164}{0.131}$ & $0.308 \err{0.029}{0.020}$ & $-0.619 \err{0.123}{0.335}$ & $-0.725 \err{0.055}{0.055}$ \\
1\%  contamination low-$z$ & $0.256 \err{0.131}{0.166}$ & $0.295 \err{0.029}{0.019}$ & $-0.852 \err{0.230}{0.455}$ & $-0.848 \err{0.065}{0.067}$ \\
5\%  contamination low-$z$ & $0.180 \err{0.158}{0.120}$ & $0.302 \err{0.029}{0.019}$ & $-0.633 \err{0.115}{0.311}$ & $-0.765 \err{0.057}{0.057}$ \\
10\% contamination low-$z$ & $0.151 \err{0.171}{0.102}$ & $0.311 \err{0.029}{0.020}$ & $-0.527 \err{0.076}{0.229}$ & $-0.678 \err{0.051}{0.049}$

\enddata

\tablecomments{BAO alone results in $\Omega_{M} = 0.293
\err{0.083}{0.077}$ and $w = -0.920 \err{0.631}{0.734}$.  Since the
input cosmology of $\Omega_{M} = 0.3$ and $w = -1$ differs slightly
from this best-fit result, there is some tension between the datasets
and their combination will cause the combined contours to move toward
the BAO-only results.}

\end{deluxetable*}

\begin{figure}
\begin{center}
\epsscale{1.25}
\rotatebox{0}{
\plotone{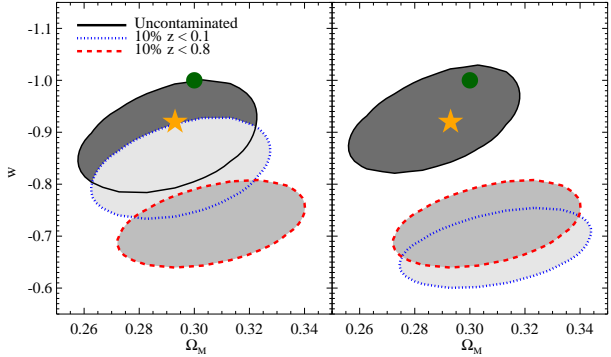}}
\caption{1$\sigma$ confidence contours in the $\Omega_{M}$--$w$ plane
for a combination of measurements of 1000 simulated SN~Ia light curves
with BAO measurements.  The black solid line and dark grey region
correspond to a sample of SNe~Ia with no SN~2006bt-like objects, while
the blue dotted line (and light grey region) and the red dashed line
(and medium grey region) corresponds to samples with 10\%
contamination of SN~2006bt-like objects in the nearby ($z < 0.1$) and
full sample, respectively.  The green dot represents the input
cosmology for the simulation ($\Omega_{M} = 0.3$ and $w = -1$), and is
offset from the most likely value of the uncontaminated sample because
of the combination with the BAO measurements, which prefer $\Omega_{M}
= 0.293$ and $w = -0.920$ (marked by the yellow star) before
combination with SN data.  The left and right panels are for
light-curve fits using the {\it flatnegav} and {\it glos} extinction
priors, respectively. [{\it See the electronic edition of the Journal
for a color version of this figure.}]\label{f:contours}}
\end{center}
\end{figure}

As \sn-like objects become a larger share of the SN sample, the
estimate of $w$ consistently increases (becoming closer to 0).  The
value derived from the uncontaminated sample is $w = -0.927
\err{0.067}{0.070}$ and $-0.893 \err{0.069}{0.072}$ for the {\it glos}
and {\it flatnegav} priors, respectively, while the the 10\%
contaminated (at all redshifts) sample yields $w = -0.725
\err{0.055}{0.055}$ and $-0.813 \err{0.065}{0.066}$ for the same
priors, respectively.  This trend is true if the \sn-like objects are
spread throughout redshift or restricted only to low redshifts, but
which of these two samples has more bias depends on the prior. For the
SN-only cosmological fits, the value of $\Omega_{M}$ is typically
biased to lower values, but only the fits with the {\it glos} prior
are substantially biased.  The 1$\sigma$ contours for the
uncontaminated and 10\% contaminated samples are shown in
Figure~\ref{f:contours}.

From our basic simulations, we see that \sn-like objects can have a
large impact on derived cosmological parameters; however, if the
fraction of \sn-like objects in the sample is low, the bias can be
much smaller than other systematic biases (if 1\% of the low-redshift
sample in contaminated by \sn-like objects, $w$ shifts by only 0.007).
We have identified one object with these peculiar properties out of
185 objects in the CfA3 sample, but we have not examined all objects
with the same scrutiny.  Within the CfA3 sample, 20\% of SNe~Ia have
measured $A_{V}$ larger than that of \sn, which is a very conservative
upper bound to the fraction of the contamination of the low-redshift
sample.


\section{Conclusions}\label{s:conc}

\sn\ is a unique object that defies the conventional wisdom of how SN~Ia
observables should correlate.  \sn\ has slowly declining light curves,
but its spectra indicate a cool photosphere.  The combination of the
broad light curves and intrinsic red colors causes light-curve fitters
to indicate that \sn\ is heavily reddened.  The object is a slight
outlier to the WLR.  Although \sn\ is underluminous for its
light-curve shape, light-curve fitters correct for this in a way that
estimates a distance modulus that is very close to the Hubble-flow
value.

We discovered the peculiar nature of \sn\ by noticing that it had a
large derived extinction value for its PGCD, but this was after
\citetalias{Hicken09:de} included \sn\ in a cosmological analysis.  We may
find similar objects by looking for comparable outliers, but it is also
possible to find \sn-like objects by examining their light curves and
spectra.  \sn\ is a clear outlier in the relationship between $\Delta
m_{15}$ and $\mathcal{R}$(Si).  Unlike SNe~Ia with similar decline
raters, \sn\ lacks a prominent second maximum in the $i$ band.  Both
of these observables could be used to find additional examples.

There are spectroscopic features which are consistent with \ion{C}{2},
but at a significantly lower velocity than other spectral features.
If these features are caused by carbon, it can be explained by a
carbon blob ejected \about 65\arcdeg\ from our line of sight.

The SN position and the host-galaxy type and spectrum indicate that
\sn\ had a very old progenitor.  If a very old progenitor is necessary
to create a \sn-like event, we may not expect to find any similar
objects at high redshift.

Contamination of \sn-like objects in samples of SNe~Ia used for
cosmological analyses is a potential source of a large bias in the
determination of $\Omega_{M}$ and $w$.  If these objects are only
\about 1\% of the low-redshift sample (similar to the sample used by
\citetalias{Hicken09:de}), then the bias should be $\lesssim 0.01$ in
$w$.  However, if these objects are a larger percentage of the sample,
this could be the dominant systematic uncertainty for SN cosmology.
It seems unlikely that \sn-like objects are more than a few percent of
the local sample, but it is much harder to characterize similar
objects at high redshift (where the rest-frame $I$ band is not
observed, and low signal-to-noise ratio spectra may not cover 
the \ion{Si}{2} $\lambda 6355$ feature).  Future studies should attempt 
to identify \sn-like objects and exclude them from cosmological analyses.

\begin{acknowledgments} 

{\it Facilities:} 
\facility{FLWO:1.5m(FAST), Keck:I(LRIS), Lick:Shane(Kast), Sloan Digital Sky Survey}

\bigskip
R.J.F.\ is supported by a Clay Fellowship, and G.N.\ is supported by
National Science Foundation (NSF) grant AST--0507475.  Supernova
research at Harvard is supported by NSF grant AST--0907903. The work
of A.V.F.'s group at U.C. Berkeley has been financed by NSF grants
AST--0607485 and AST--0908886, as well as by the TABASGO Foundation.
We are especially grateful to S.\ Blondin who has tirelessly
maintained the CfA spectroscopic database and reduced several of the
spectra presented in this paper; without his efforts, this study would
not have been possible.  We gratefully acknowledge W.\ High, A.\
Howell, S.\ Jha, D.\ Kasen, A.\ Rest, and R.\ Thomas for discussing
this interesting object with us.  W.\ Brown, M.\ Calkins, T.\ Groner,
M.\ Moore, W.\ Peters, and D.\ Wong helped obtain the spectra
presented in this paper; we thank them for their time.  We thank the
referee, D.\ Branch, who provided insightful comments.  We are
indebted to the staffs at the Lick, Keck, and Fred L.\ Whipple
Observatories for their dedicated services.

Some of the data presented herein were obtained at the W.~M.\ Keck
Observatory, which is operated as a scientific partnership among the
California Institute of Technology, the University of California, and
the National Aeronautics and Space Administration (NASA); the
observatory was made possible by the generous financial support of the
W.~M.\ Keck Foundation.  This research has made use of the NASA/IPAC
Extragalactic Database (NED), which is operated by the Jet Propulsion
Laboratory, California Institute of Technology, under contract with
NASA.

We made extensive use of the Hydra Cluster administered by the
Computation Facility of the CfA and the Odyssey Cluster administered
by the FAS-IT Research Computing Group.  We are grateful to the staff
that maintains these facilities.

Funding for the SDSS and SDSS-II has been provided by the Alfred P.\
Sloan Foundation, the Participating Institutions, the National Science
Foundation, the U.S. Department of Energy, the National Aeronautics
and Space Administration, the Japanese Monbukagakusho, the Max Planck
Society, and the Higher Education Funding Council for England. The
SDSS Web Site is http://www.sdss.org/.

The SDSS is managed by the Astrophysical Research Consortium for the
Participating Institutions. The Participating Institutions are the
American Museum of Natural History, Astrophysical Institute Potsdam,
University of Basel, University of Cambridge, Case Western Reserve
University, University of Chicago, Drexel University, Fermilab, the
Institute for Advanced Study, the Japan Participation Group, Johns
Hopkins University, the Joint Institute for Nuclear Astrophysics, the
Kavli Institute for Particle Astrophysics and Cosmology, the Korean
Scientist Group, the Chinese Academy of Sciences (LAMOST), Los Alamos
National Laboratory, the Max-Planck-Institute for Astronomy (MPIA),
the Max-Planck-Institute for Astrophysics (MPA), New Mexico State
University, Ohio State University, University of Pittsburgh,
University of Portsmouth, Princeton University, the United States
Naval Observatory, and the University of Washington.

\end{acknowledgments}

\bibliographystyle{fapj}
\bibliography{astro_refs}


\end{document}